\documentclass[
 reprint,amsmath,amssymb,aps,
prl,
]{revtex4-2}

\usepackage{graphicx}
\usepackage{dcolumn}
\usepackage{bm}
\usepackage{hyperref}
\hypersetup{colorlinks, linkcolor={blue}, citecolor={blue}, urlcolor={blue}}
\usepackage{xcolor}
\usepackage{pifont}

\begin{document}
\title{Critically Enhanced Magnon Transport in Low-dimensional Magnets}

\author{Ping Tang}
\email{tang.ping.a2@tohoku.ac.jp}
\affiliation{Institute for Materials Research, Tohoku University,
2-1-1 Katahira, Sendai 980-8577, Japan} 
\date{\today}
\begin{abstract}
Transport properties of (quasi)particles in condensed matter depend profoundly on the spatial dimension. Motivated by recent advances in growing ultrathin magnetic films and monolayer van der Waals magnets, we present a theory of magnon transport in magnetic films spanning the crossover from bulk to the two-dimensional (2D) limit. We find a magnon conductivity that diverges~\emph{logarithmically} in magnetically soft but stable (quasi)2D magnets with long-range dipolar interactions. This critical enhancement is absent in bulk systems and may explain the unusually large magnon conductivities recently observed in ultrathin yttrium iron garnet films. Our results reveal an intrinsic mechanism for enhanced magnon transport in low dimensions and highlight the potential for engineering high-efficiency magnon conductors in atomically thin magnets.
\end{abstract}

\maketitle
\emph{Introduction---}Transport phenomena lie at the heart of nonequilibrium condensed matter physics, and their behavior depends critically, and often nonperturbatively, on spatial dimensionality. A prominent example is Anderson localization, i.e., the localization of electrons in 1D and 2D systems in the presence of arbitrarily weak disorder~\cite{kramer1993localization}. Spin transport, a central theme in spintronics, has been extensively investigated in bulk materials. When mediated by mobile electrons, it suffers from Joule heating and relatively short spin diffusion lengths. Magnons, the collective excitations of magnetic order, offer an attractive charge-free alternative capable of transporting spin angular momentum over long distances~\cite{chumak2015magnon}. In magnetic insulators, magnons can be electrically injected by the spin Hall effect in heavy-metal contacts and detected nonlocally as an electric voltage in the other contact through the inverse spin Hall effect~\cite{kajiwara2010transmission, PhysRevLett.109.096603, cornelissen2015long, cornelissen2016magnon,PhysRevB.93.060403,li2016observation}. However, even in record low-damping yttrium iron garnet (YIG), the magnon conductivity, a key material parameter that quantifies the efficiency of magnon-mediated spin transport, typically does not exceed the electronic conductivity of bad metals~\cite{cornelissen2015long, cornelissen2016magnon}.

Atomically thin or (quasi)2D magnets are attractive for highly integrated data storage and processing device~\cite{burch2018magnetism, gong2019two,cortie2020two}, while their intrinsically enhanced spin fluctuations~\cite{gibertini2019magnetic} may also enable new functionalities such as a reduced critical spin torque for magnetization switching \cite{PhysRevB.103.094442}. According to the Mermin-Wagner-Hohenberg theorem \cite{PhysRevLett.17.1133,PhysRev.158.383}, an isotropic short-range exchange interaction alone cannot sustain long-range magnetic order in 2D, since the occupation of a gapless (Goldstone) magnon diverges at any finite temperature. Magnetism can exist in (quasi-)2D systems \cite{RevModPhys.72.225} only in the presence of magnetic anisotropies supplied by spin-orbit coupling or \emph{long-range} dipole-dipole interactions \cite{doring1961sattigungsmagnetisierung,PhysRev.130.2223,davis1963effect}. The former suppresses thermal spin fluctuations 
 by ``gapping" the magnon spectrum, as occurs in Ising-type magnetic monolayers, such as FePS$_3$~\cite{lee2016ising}, CrI$_3$~\cite{huang2017layer,lado2017origin,xu2018interplay,pizzochero2020magnetic, kim2019exploitable, soriano2020magnetic}, and Fe$_3$GeTe$_2$ \cite{fei2018two, deng2018gate}. In contrast, the long-range magneto-dipolar interaction, which typically plays a minor role in bulk materials, stabilizes 2D magnetic order by \emph{qualitatively} reshaping the magnon dispersion \cite{maleev1976dipole, PhysRevB.33.6519}. This may explain the robust magnetism in (quasi-)2D \emph{soft} magnetic films with negligible magnetocrystalline (spin-orbit-induced) anisotropy, such as YIG~\cite{dorsey1993epitaxial,PhysRevMaterials.3.034403} and van der Waals compound Cr$_2$Ge$_2$Te$_6$ \cite{gong2017discovery}.

Enhanced spin fluctuations in low dimensions, on the other hand, imply higher magnon conductivities, opening the prospect of tuning device parameters by dimensional engineering. Recent experiments on spin transport increasingly address van der Waals magnets \cite{PhysRevX.9.011026,PhysRevB.101.205407,PhysRevB.107.L180403} and ultrathin magnetic films~\cite{PhysRevB.103.214425, li2022anisotropic, wei2022giant}.~Remarkably, Wei \textit{et al.}~\cite{wei2022giant} reported dramatically enhanced magnon conductivities in few-unit-cell-thick YIG films, in sharp contrast to the Fuchs-Sondheimer model for electronic conductivity in metallic films~\cite{fuchs1938conductivity, sondheimer2001mean}, in which decreasing thickness reduces conductivity due to surface scattering. Wei \textit{et al.}~\cite{wei2022giant} explain their observations by the high mobility of low-frequency magnons that carry a higher spectral weight close to the band edge in one or two than in three dimensions. Here we revisit this issue because that argument is based on a purely exchange-coupled 2D magnet with constant magnon density of states, in which the infrared divergence of the magnon number without the dipolar interaction reflects the instability of the magnetic order noted above. 


In this Letter, we present a microscopic theory of magnon transport in magnetic films across the thickness crossover from the bulk to the quasi-2D limit, incorporating all essential magnetic interactions, including exchange, long-range dipolar interactions, and magnetocrystalline anisotropy. We find that the magnon conductivity logarithmically diverges in ultrasoft magnets as the film thickness reduces to the quasi-2D regime, a behavior directly relevant to the ultrathin YIG films studied in Ref.~\cite{wei2022giant}. This critical enhancement originates from the dipolar-modified magnon dispersion, which simultaneously removes the infrared divergence in the thermally excited magnon population. The high magnon conductivity in the quasi-2D regime remains robust against defect and phonon scattering, underscoring the promise of (quasi-)2D magnets as efficient spintronic and magnonic platforms. We further derive a critical scaling relation for the magnon conductivity as a function of the film thickness and magnetocrystalline anisotropy, analogous to a phase transition but for a non-equilibrium transport property.

\begin{figure}
    \centering
    \includegraphics[width=8.6cm]{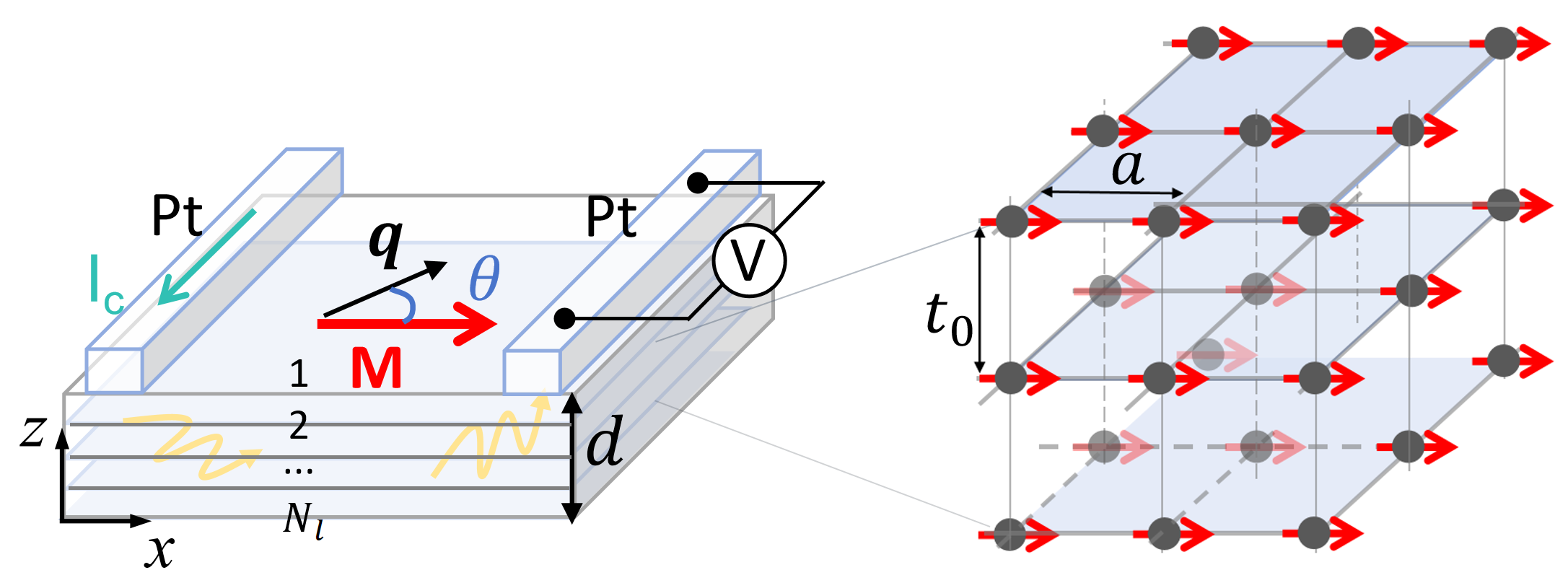}
    \caption{Schematics of magnon transport in a magnetic film of $N_{l}$ atomic layers.}
    \label{Fig-schematic}
\end{figure}

\emph{Model---}We consider a magnetic film of $N_{l}$ atomic layers within $xy$ plane, with the equilibrium magnetization along $x$-direction [Fig.~{\ref{Fig-schematic}}]. The spin Hamiltonian reads
\begin{align}
    \hat{\mathcal{H}}=&-J\sum_{\langle ij\rangle} \hat{\mathbf{S}}_{i}\cdot\hat{\mathbf{S}}_{j}-\frac{K}{2}\sum_{i}(\hat{S}_{i}^{x})^2-\gamma\hbar H\sum_{i}\hat{S}_{i}^{x}\nonumber\\
    &+\frac{(\gamma\hbar)^2}{2}\sum_{i\neq j}\left[\frac{\hat{\mathbf{S}}_{i}\cdot\hat{\mathbf{S}}_{j}}{\vert \mathbf{r}_{ij}\vert^3}-\frac{3(\mathbf{r}_{ij}\cdot\hat{\mathbf{S}}_{i})(\mathbf{r}_{ij}\cdot\hat{\mathbf{S}}_{j})}{\vert \mathbf{r}_{ij}\vert^5}\right],\label{Ham}
\end{align}
 where $J$ is the nearest-neighbor exchange coupling, $K$ is the on-site magnetocrystalline anisotropy (without shape anisotropy), and $H$ is an external magnetic field that specifies the magnetization direction in the easy ($x$) axis. The last term represents the longe-range magnetodipolar interaction between spins at sites $i$ and $j$ separated by $\mathbf{r}_{ij}=\mathbf{r}_{j}-\mathbf{r}_{i}$. In soft magnets such as YIG, where the magnetocrystalline anisotropy is very weak, dipolar interactions primarily stabilize the in-plane magnetic order in the 2D limit in the absence of an external magnetic field \cite{PhysRevB.43.6015}. 

Within the truncated Holstein-Primakoff transformation, the spin operators are expressed as $\hat{\mathbf{S}}_{i}\simeq(S-\hat{a}_{i}^{\dagger}\hat{a}_{i})\mathbf{x}+\sqrt{S/2}[(\mathbf{y}-i\mathbf{z})\hat{a}_{i}+\text{H.c.}]$ in terms of magnon creation and annihilation operators $\hat{a}_{i}^{(\dagger)}$ at site $i$. For an extended film, 
\begin{align}
\hat{a}_{i}\left(\boldsymbol{\rho}_{l}, z_{l}\right)=\frac{1}{\sqrt{N_{a}}}\sum_{\mathbf{q}}\hat{a}_{l,\mathbf{q}} e^{i\mathbf{q}\cdot\boldsymbol{\rho}_{l}},\label{magnon}
\end{align}
where $N_{a}$ is the number of sites per atomic layer, $l$ labels the layer index,  $\mathbf{q}=(q_{x}, q_{y})$ is an in-plane wave vector, and $\mathbf{r}_{i}=(\boldsymbol{\rho}_{l}, z_{l})$ denotes the position of site $i$. $\hat{a}_{l,\mathbf{q}}^{(\dagger)}$ annihilates (creates) a magnon with wave vector $\mathbf{q}$ in layer $l$. Assuming a simple cubic lattice [Fig.~\ref{Fig-schematic}], we arrive at the $2N_l\times 2N_l$ quadratic magnon Hamiltonian
\begin{align}
\hat{\mathcal{H}}=\frac{\gamma\hbar}{2}\sum_{\mathbf{q}}\left( \begin{matrix}
 \hat{\mathbf{a}}_{\mathbf{q}}^{\dagger}\\
 \hat{\mathbf{a}}_{-\mathbf{q}}
\end{matrix} \right)^{T}\left(\begin{matrix}
 \hat{\mathcal{A}}_{\mathbf{q}} & \hat{\mathcal{B}}_{\mathbf{q}}\\
 \hat{\mathcal{B}}_{-\mathbf{q}}^{\ast} & \hat{\mathcal{A}}_{-\mathbf{q}}^{\ast}
\end{matrix}\right)\left( \begin{matrix}
 \hat{\mathbf{a}}_{\mathbf{q}}\\
 \hat{\mathbf{a}}_{-\mathbf{q}}^{\dagger}
\end{matrix} \right)\label{mHam}
\end{align}
where $\hat{\mathbf{a}}_{\mathbf{q}}=(\hat{a}_{1,\mathbf{q}}, \cdots, \hat{a}_{N_l,\mathbf{q}})^{T}$ is an $N_l$-dimensional column vector of layer-resolved magnon operators, and the elements of the $N_l\times N_l$ matrices $\hat{\mathcal A}_{\mathbf q}$ and $\hat{\mathcal B}_{\mathbf q}$ are
 \begin{align}
[\hat{\mathcal{A}}_{\mathbf{q}}]_{ll^{\prime}}=&\hbar(\omega_{l}+\frac{1}{2}\omega_{M}) \delta_{l,l^{\prime}}-JS\delta_{l\pm1,l^{\prime}}\nonumber\\
&-\frac{\hbar\omega_{M}}{4}qt_{0} e^{-q\vert z_{ll^{\prime}}\vert}\cos^2\theta,\\
[\hat{\mathcal{B}}_{\mathbf{q}}]_{ll^{\prime}}=&-\frac{\hbar\omega_{M}}{2} \delta_{l,l^{\prime}}+\frac{\hbar\omega}{2}qt_{0} e^{-q\vert z_{ll^{\prime}}\vert}\nonumber\\&\times\left\{\frac{1}{2}(1+\sin^2\theta)-\text{sgn}(z_{ll^{\prime}})\sin\theta\right\}. \label{Bq}
\end{align}   
Here, $\hbar\omega_{l}=2JS [2-\cos (q_{x}a)-\cos (q_{y}a)]+KS+\gamma\hbar H+\zeta_{l}JS$ is a single-layer magnon energy, where $\zeta_{l}$ is the number of nearest-neighbor layers of layer $l$, $\omega_{M}=4\pi \gamma M_{s}$ measures the dipolar interaction strength, $t_{0}$ is the interlayer spacing (equal to the in-plane lattice constant $a$ for the simple cubic lattice), and $z_{ll^{\prime}}=z_{l}-z_{l^{\prime}}$. The sign function satisfies $\mathrm{sgn}(z_{ll'})=\pm1$ for $z_{ll'}\gtrless0$ and $\mathrm{sgn}(z_{ll'})=0$ for $z_{ll'}=0$. $q=\sqrt{q_{x}^2+q_{y}^2}$, and $\theta=\tan^{-1}q_{y}/q_x$ is the angle of $\mathbf{q}$ relative to the magnetization ($x$) direction. $\zeta_{l}=1$ for the two outermost layers, while $\zeta_{l}=2$ for interior layers. In van der Waals antiferromagnets with atomically flat surfaces, it has been shown that the layer-dependent coordination number $\zeta_l$ gives rise to nanoscale-confined exchange surface spin waves~\cite{bchp-xqtn}. Unless otherwise specified, we assume $\zeta_{l}=2$, disregarding the distinction between the outermost and interior layers for a realistic film with surface roughness. Eq.~(\ref{mHam}) can be diagonalized via the Bogoliubov transformation $(\hat{\mathbf{a}}_{\mathbf{q}},\hat{\mathbf{a}}_{-\mathbf{q}}^{\dagger})^{T}=\hat{T}_{\mathbf{q}}(\hat{\mathbf{b}}_{\mathbf{q}}, \hat{\mathbf{b}}_{-\mathbf{q}}^{\dagger})^{T}$, yielding
\begin{equation}
\hat{\mathcal{H}}=\sum_{n,\mathbf{q}} \hbar\omega_{n,\mathbf{q}}\left(\hat{b}_{n,\mathbf{q}}^{\dagger}b_{n,\mathbf{q}}+\frac{1}{2}\right).
\end{equation}
where $\omega_{n,\mathbf{q}}$ is the magnon dispersion with band index $n$ (not to be confused with the layer index $l$), and $\hat{b}_{n,\mathbf{q}}^{(\dagger)}$ annihilates (creates) a magnon eigenmode. The transformation matrix $\hat{T}_{\mathbf{q}}$ is para-unitary and satisfies $\hat{T}_{\mathbf{q}}\hat{\eta}\hat{T}_{\mathbf{q}}^{\dagger}=\hat{T}_{\mathbf{q}}^{\dagger}\hat{\eta
}\hat{T}_{\mathbf{q}}=\hat{\eta}$ with $\hat{I}_{N_l\times N_l}$ being the $N_l\times N_l$ identity matrix. Because of the dipolar interaction, the magnon dispersion $\omega_{n,\mathbf{q}}$ is intrinsically nonreciprocal under $q_y\rightarrow -q_y$, as implied by the term proportional to $\sin\theta$ in Eq.~(\ref{Bq}).

\emph{Magnon transport---}We now address magnon transport in the nonlocal geometry~\cite{ cornelissen2015long}, in which the film is magnetized in the plane along the $x$ direction while magnons are electrically injected and detected by the spin Hall and inverse spin Hall effect in two heavy-metal strips parallel to the $y$ axis, as illustrated in Fig.~~\ref{Fig-schematic}. The gradient in the non-equilibrium magnon accumulation \(\mu_m\) excited by the injector strip drives a lateral diffusion of magnons from the injector to the detector contact. Here we focus on films with thicknesses much smaller than the magnon diffusion length, disregarding diffusion in the thickness ($z$) direction. The magnon current density, $ j_{m}=-(\hbar/e^2)\sigma_{m} \partial_{x}\mu_m $, defines the magnon conductivity $\sigma_{m}$ in units of the conventional electronic conductivity S/m. The semiclassical Boltzmann equation within the relaxation-time approximation leads to
\begin{align}
\sigma_{m}=\frac{-e^{2}}{\hbar d}\int\frac{d^2\mathbf{q}}{(2\pi)^2}\sum_{n}\tau_{n,\mathbf{q}}\left(\frac{\partial\omega_{n,\mathbf{q}}}{\partial q_{x}}\right)^{2}\frac{\partial n_{P}\left(\omega_{n,\mathbf{q}}\right)}{\partial \omega_{n,\mathbf{q}}}\label{Conduct}
\end{align}  
where $n_{P}(\omega_{n,\mathbf{q}})=[\exp(\hbar\omega_{n,\mathbf{q}}/k_{B}T)-1]^{-1}$ is the Planck distribution at temperature \(T\), $d=N_{l} t_{0}$ is the film thickness, and $\partial\omega_{n,\mathbf{q}}/\partial q_{x}$ is the magnon group velocity in the transport ($x$) direction. The relaxation times $\tau_{n,\mathbf{q}}$ are dominated by scattering from impurities, phonons, surface roughness, and magnon--magnon interactions. In ultrathin films, they may differ from that of the magnons in bulk magnets~\cite{PhysRevB.96.174416,PhysRevB.99.184442}. Coherent parametric pumping experiments \cite{PhysRevB.97.214405} report a nearly constant relaxation rate up to a wave vector $6\times10^5\,$rad/cm for magnons in the lowest subband (\(n=0\)) of an ultrathin YIG film. Refs.~\cite{boona2014magnon, PhysRevB.92.064413, PhysRevB.92.054436} conclude from the observation of a strong magnetic field dependence at room temperature that magnon transport in YIG is dominated by low-energy (subthermal) magnons. The detailed dependence of $\tau_{n,\mathbf{q}}$ on the subband structure, sample shape, crystalline quality, temperature, etc., is beyond the scope of our paper. Here we adopt in Eq.~(\ref{Conduct}) a ($n,\mathbf{q}$)-independent relaxation time $\tau$, focusing on the intrinsic mechanism for the enhanced magnon transport in low dimensions. 

\begin{figure}
    \centering
    \includegraphics[width=8.6cm]{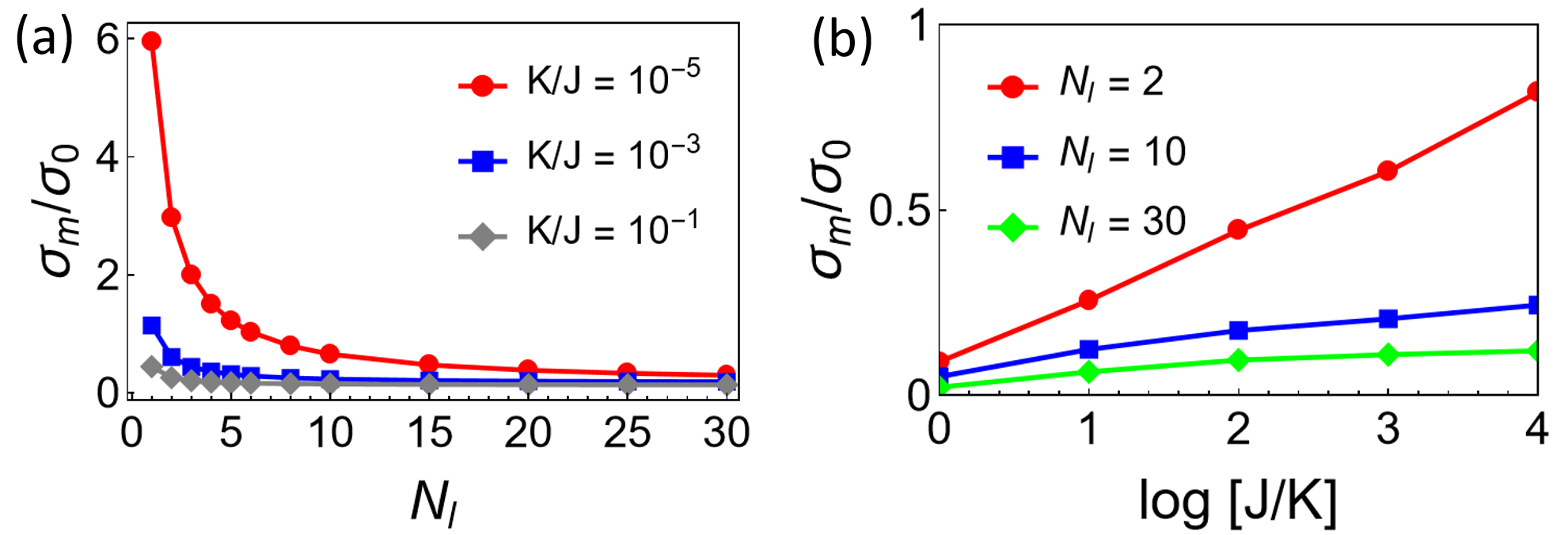}
    \caption{Magnon conductivity at room temperature as a function of the atomic layer number $N_{l}$ and the magnetocrystalline anisotropy $K$, normalized by $\sigma_{0}=\tau k_{B}T/(a\hbar^{2})$. The exchange interaction and magnetization are chosen to be close to those of YIG with $M=1.94\times 10^5\,$A/m and $J=1.64\,$meV~\cite{rezende2020fundamentals}, while the anisotropy constant \(K\) is a free parameter. The external magnetic field is set to zero.} \label{Fig-conductivity}
\end{figure}

Fig.~\ref{Fig-conductivity} presents the calculated room-temperature magnon conductivity $\sigma_{m}$ (at $H=0$) as a function of the anisotropy-to-exchange ratio $K/J$ and the number of atomic layers $N_{l}$ (film thickness). We find that $\sigma_{m}$ increases as either $K/J$ or $N_{l}$ is reduced. Remarkably, for ultrathin films with $N_{l}\lesssim10$, $\sigma_{m}$ exhibits a pronounced linear scaling with $\text{log}(J/K)$, implying a logarithmic divergence in the limit of vanishing magnetocrystalline anisotropy. This behavior accounts for the dramatic enhancement of the magnon conductivity observed in magnetically soft YIG films as the thickness is reduced to the quasi-2D regime~\cite{wei2022giant}.

\emph{Critical scaling relation---}We now derive the critical behavior of the quasi-2D magnon conductivity as the magnetocrystalline anisotropy decreases. In ultrathin films with $k_{B}T\lesssim JS (a\pi/d)^{2}$, where the thermal energy is insufficient to excite spatially nonuniform standing-wave modes along the thickness direction, we may consider in Eq.~(\ref{Conduct}) only the contribution from the lowest magnon subband ($n=0$), whose amplitude is uniform across different layers~\cite{notelowest}. Although this quasi-2D condition is well satisfied in thin films of a few atomic layers, the inclusion of higher standing-wave modes does not alter the scaling relation derived below, which is governed by long-wavelength magnons in the lowest mode. Then $\hat{a}_{i}(\boldsymbol{\rho}_{l},z_{l})\simeq (N_{l}N_{a})^{-1/2}\sum_{\mathbf{q}}\hat{a}_{\mathbf{q}}e^{i\mathbf{q}\cdot\boldsymbol{\rho}_{l}}$, which reduces Eq.~(\ref{mHam}) to the harmonic oscillator form
\begin{align}
\hat{\mathcal{H}}\simeq&\sum_{\mathbf{q}}\mathcal{A}_{\mathbf{q}}\hat{a}_{\mathbf{q}}^{\dagger}\hat{a}_{\mathbf{q}}+\frac{1}{2}(\mathcal{B}_{\mathbf{q}}\hat{a}_{\mathbf{q}}\hat{a}_{-\mathbf{q}}+\text{H.c.})\nonumber\\=&\sum_{\mathbf{q}}\hbar\omega_{\mathbf{q}}\left(\hat{b}_{\mathbf{q}}^{\dagger}\hat{b}_{\mathbf{q}}+\frac{1}{2}\right) \label{umH}
\end{align}
where $\hat{a}_{\mathbf{q}}=\sqrt{N_{l}}\hat{a}_{l,\mathbf{q}}$ annihilates a magnon in the lowest subband \(n=0\) state. The coefficients $\mathcal{A}_{\mathbf{q}}\equiv N_{l}^{-1}\sum_{ll^{\prime}}[\hat{\mathcal{A}}_{\mathbf{q}}]_{ll^{\prime}}=JSa^2q^2+KS+\gamma\hbar H+\frac{1}{2}\hbar\omega_{M}[1+(F_{q}-1)\cos^2\theta]$ and $\mathcal{B}_{\mathbf{k}}\equiv N_{l}^{-1}\sum_{ll^{\prime}}[\hat{\mathcal{B}}_{\mathbf{k}}]_{ll^{\prime}}=\frac{1}{2}\hbar\omega_{M}[\sin^2\theta -(1+\sin^2\theta)F_{q}]$ contain the dipolar form factor $F_{q}=1-\frac{qa}{2N_{l}}\sum_{ll^{\prime}}e^{-q\vert z_{ll^{\prime}}\vert}$. 
The Bogoliubov transformation $\hat{a}_{\mathbf{q}}=u_{\mathbf{q}}\hat{b}_{\mathbf{q}}-v_{\mathbf{q}}b_{-\mathbf{q}}^{\dagger}$ diagonalizes Eq.~(\ref{umH}) and yields the dispersion of the uniform band mode $\hbar\omega_{\mathbf{q}}=\sqrt{\mathcal{A}_{\mathbf{q}}^2-\mathcal{B}_{\mathbf{q}}^2}$, with $u_{\mathbf{q}}=\sqrt{(\mathcal{A}_{\mathbf{q}}+\hbar\omega_{\mathbf{q}})/(2\hbar\omega_{\mathbf{q}})}$ and $v_{\mathbf{q}}=\sqrt{(\mathcal{A}_{\mathbf{q}}-\hbar\omega_{\mathbf{q}})/(2\hbar\omega_{\mathbf{q}})}$. In the long-wavelength regime, $F_{q}=1-qd/2+\mathcal{O}(q^2)$ and \begin{align}
\omega_{\mathbf{q}}=\sqrt{\Delta^2+ \alpha(\theta) qd+\mathcal{O}(q^2)},\label{energy}
\end{align} 
where $\Delta=\gamma\sqrt{H_{x}(H_{x}+4\pi M)}$ is the magnon gap induced by the magnetocrystalline anisotropy and/or an external magnetic field, with $H_{x}=KS/(\gamma\hbar)+H$ the effective field along the $x$ direction, and $\alpha(\theta)=2\pi \gamma^2 M[(H_{x}+4\pi M)\sin^2\theta-H_{x}]$ characterizes the emergent \emph{linear} dispersion from the dipolar interaction. In the limit $\Delta\rightarrow 0$, Eq.~(\ref{energy}) reduces to the 2D magnon dispersion with $d\rightarrow a$~\cite{PhysRevB.33.6519, PhysRevB.43.6015}. Note that even though the magnon spectrum becomes gapless when $\Delta\rightarrow 0$, the dipolar interaction regularizes the infrared divergence of the magnon population by reshaping the exchange-induced parabolic dispersion into $\sim\sqrt{q}$, thereby stabilizing 2D magnetic order~\cite{PhysRevB.33.6519}; however, the long-wavelength magnons in the lowest band generate an infrared divergence in the magnon conductivity, as shown below.

Since long-wavelength magnons with $\hbar\omega_{\mathbf{q}}\lesssim k_{B}T$ cause the divergent conductivity in the limit of a vanishing anisotropy, we may approximate the Planck distribution in Eq.~(\ref{Conduct}) by its classical limit $n_{P}(\omega_{\mathbf{q}})\simeq  k_{B}T/\hbar\omega_{\mathbf{q}}$. With an ultraviolet cut-off $\Lambda$ for the in-plane wave vector~\cite{notecutoff}, we find in the limit $\Delta\ll \omega_{M}$ 
\begin{equation}
\sigma_{m}\left(\Delta\right)\simeq\sigma_\Lambda+\frac{\tau e^2k_{B}T}{4\hbar^2 d}\ln \frac{\omega_{M}}{\Delta},\label{scaling}
\end{equation}
where $\sigma_{\Lambda}$ is  a cutoff-dependent contribution that does not depend critically on $\Delta$,
\begin{equation}
\sigma_{\Lambda}=\frac{\tau e^2 k_{B}T}{8\hbar^2d}\left\{[\text{ln}(\Lambda d)-1]+\int \frac{d\theta}{\pi} \cos^2\theta \, \, \, \text{ln}\frac{\sin^2\theta}{2}\right\}\nonumber.
\end{equation}
Since $\Delta\ll\omega_M$, $\sigma_m$ is dominated by the second, $\Lambda$-independent term in Eq.~(\ref{scaling}) that diverges
\emph{logarithmically} as $\Delta\rightarrow0$, while its thickness dependence is governed by the prefactor $\tau k_{B}T/d$. In thin films, the magnon relaxation time becomes thickness dependent since surface scattering introduces an additional relaxation rate that scales as $\beta/d$, yielding $\tau^{-1}=\tau_{\text{3D}}^{-1}+\beta/d$, where $\tau_{\text{3D}}^{-1}$ is the intrinsic bulk relaxation rate and $\beta$ a constant characterizing surface scattering. This additional boundary scattering channel leads to $\sigma_{m}\sim k_{B}T \tau_{\text{3D}}/(\beta\tau_{\text{3D}}+d)\ln(\omega_{M}/\Delta)$, which softens but does not remove the dimensional enhancement of magnon conductivity in the ultrathin limit, in sharp contrast to electronic conductivity in metallic films~\cite{fuchs1938conductivity, sondheimer2001mean}.

\emph{Conclusions---}We present a microscopic theory of magnon transport in magnetic films spanning the crossover from bulk to the quasi-2D limit, incorporating the exchange interactions, magnetocrystalline anisotropy, and long-range dipolar interactions. We find that reducing the dimensionality substantially enhances the magnon conductivity by spin fluctuations that are amplified compared with the bulk case. In the quasi-2D limit and vanishing spin wave gap, the long-range dipolar interaction both stabilizes the magnetic order and causes a logarithmically divergent magnon conductivity. This behavior is consistent with the large magnon conductivities observed in magnetically soft, ultrathin YIG films with nearly constant magnetization down to three atomic layers~\cite{wei2022giant}. We predict a similar critical enhancement for (quasi-)2D van der Waals antiferromagnets with out-of-plane magnetization that become magnetically soft at the spin-flip transition~\cite{PhysRevB.107.L180403}. Our results herald the potential of atomically thin magnets for scalable magnon-based spintronic devices. 

\emph{Acknowledge.---}The author thanks Gerrit E. W. Bauer for useful suggestions on the manuscript. This work was supported by JSPS KAKENHI Grant-in-Aid for Scientific Research (B), Grant No. 26K00625.

\bibliography{reference}

\end{document}